\documentclass[aps,showpacs,pre,twocolumn,superscriptaddress,floatfix]{revtex4}
 \usepackage{graphicx,bm,amssymb,amsmath,dcolumn,hyperref}
 \usepackage{multirow}
 \usepackage{enumerate}
 \usepackage{color}
 \usepackage{subfigure}  
 \usepackage{epstopdf}
 \usepackage{bbm}
 \usepackage{graphicx}
 \usepackage{hyperref}
 \usepackage{threeparttable}
 \usepackage{amsthm}
 \usepackage{mathtools}

\begin{document}
\newcommand{\be}{\begin{equation}}
\newcommand{\ee}{\end{equation}}
\newcommand{\half}{\frac{1}{2}}
\newcommand{\ith}{^{(i)}}
\newcommand{\im}{^{(i-1)}}
\newcommand{\gae}
{\,\hbox{\lower0.5ex\hbox{$\sim$}\llap{\raise0.5ex\hbox{$>$}}}\,}
\newcommand{\lae}
{\,\hbox{\lower0.5ex\hbox{$\sim$}\llap{\raise0.5ex\hbox{$<$}}}\,}

\definecolor{blue}{rgb}{0,0,1}
\definecolor{red}{rgb}{1,0,0}
\definecolor{green}{rgb}{0,1,0}
\newcommand{\blue}[1]{\textcolor{blue}{#1}}
\newcommand{\red}[1]{\textcolor{red}{#1}}
\newcommand{\green}[1]{\textcolor{green}{#1}}

\newcommand{\scrA}{{\mathcal A}}
\newcommand{\scrE}{{\mathcal E}} 
\newcommand{\scrF}{{\mathcal F}} 
\newcommand{\scrL}{{\mathcal L}}
\newcommand{\scrM}{{\mathcal M}} 
\newcommand{\scrN}{{\mathcal N}}
\newcommand{\scrS}{{\mathcal S}}
\newcommand{\scrs}{{\mathcal s}}
\newcommand{\scrP}{{\mathcal P}}
\newcommand{\scrO}{{\mathcal O}}
\newcommand{\scrR}{{\mathcal R}}
\newcommand{\scrC}{{\mathcal C}}
\newcommand{\scrV}{{\mathcal V}}
\newcommand{\scrD}{{\mathcal D}}
\newcommand{\dm}{d_{\rm min}}
\newcommand{\rhojunction}{\rho_{\rm j}}
\newcommand{\rhojunctionLim}{\rho_{{\rm j},0}}
\newcommand{\rhobranch}{\rho_{\rm b}}
\newcommand{\rhobranchLim}{\rho_{{\rm b},0}}
\newcommand{\rhononbridge}{\rho_{\rm n}}
\newcommand{\rhononbridgeLim}{\rho_{{\rm n},0}}
\newcommand{\percolationCluster}{C}
\newcommand{\leafFreeCluster}{C_{\rm \ell f}}
\newcommand{\bridgeFreeCluster}{C_{\rm bf}}


\newcommand{\cartoonfFig}{\includegraphics[scale=0.24]{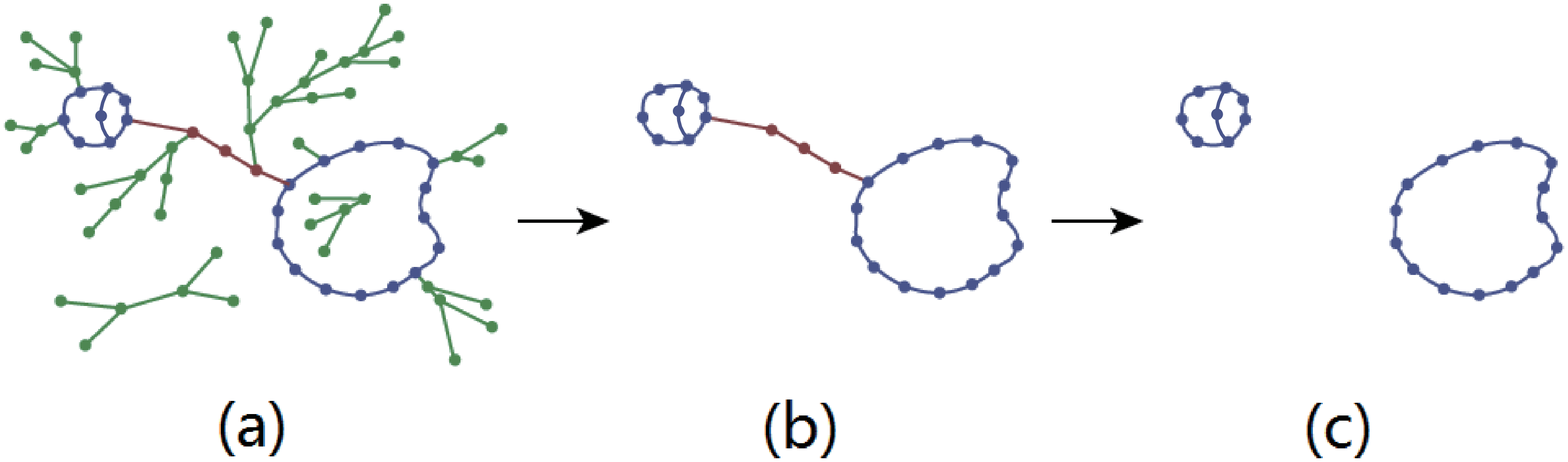}}
\newcommand{\bondensiFig}{\includegraphics[scale=1.1]{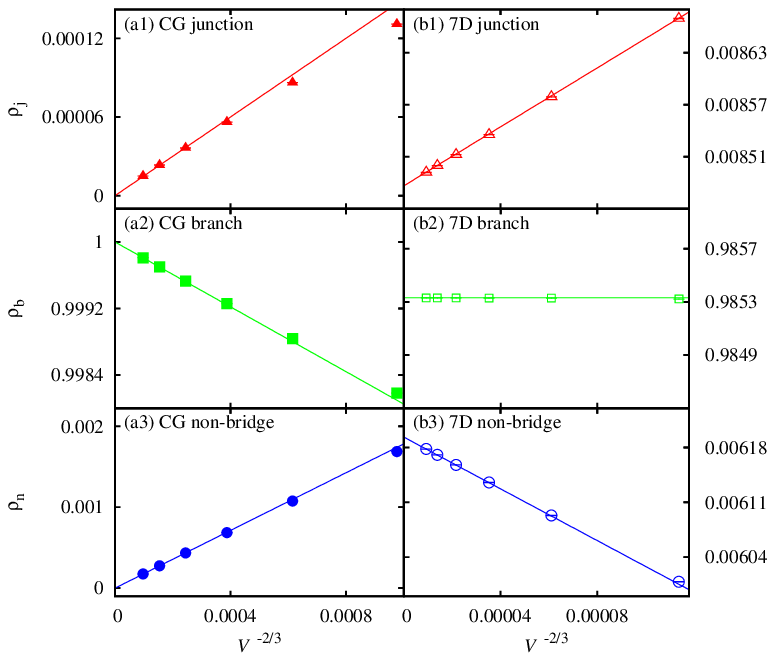}}
\newcommand{\fractalfFig}{\includegraphics[scale=1.3]{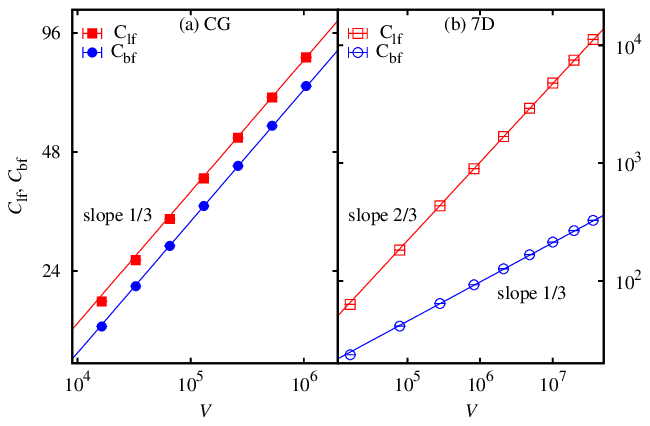}}
\newcommand{\clusternFig}{\includegraphics[scale=1.1]{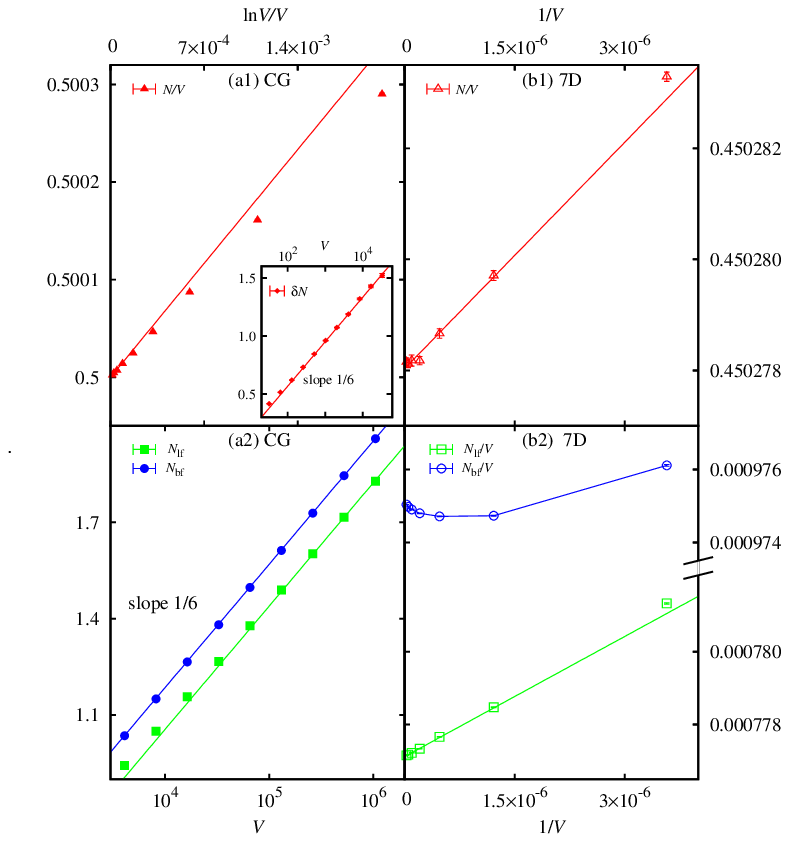}}
\newcommand{\distribpFig}{\includegraphics[scale=0.9]{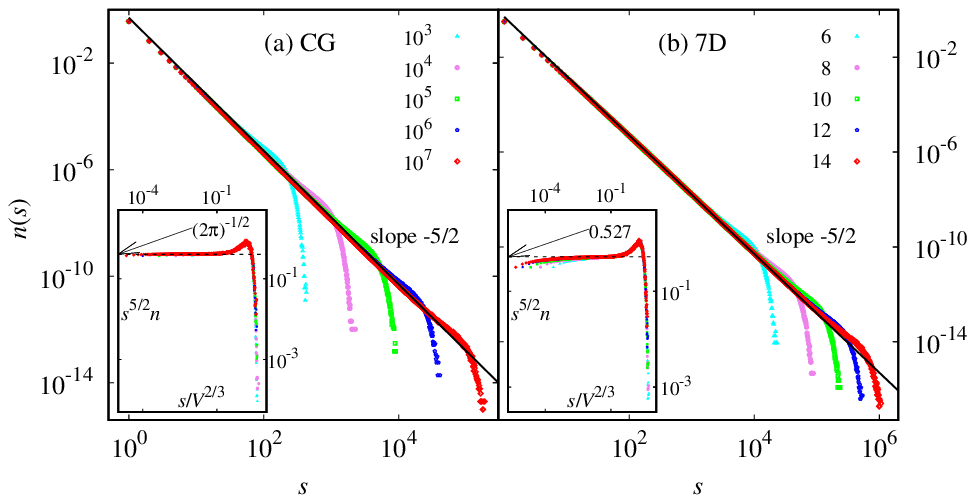}}
\newcommand{\metricfaFig}{\includegraphics[scale=0.9]{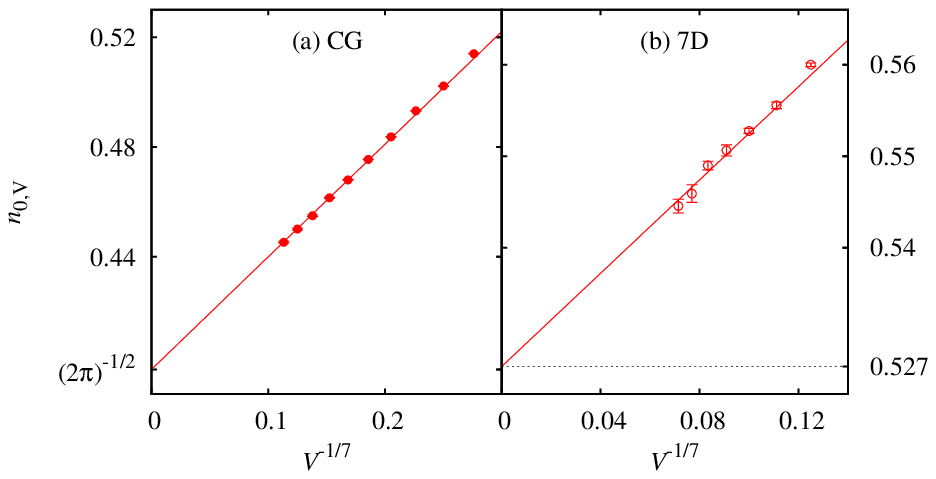}}
\newcommand{\distribfFig}{\includegraphics[scale=0.9]{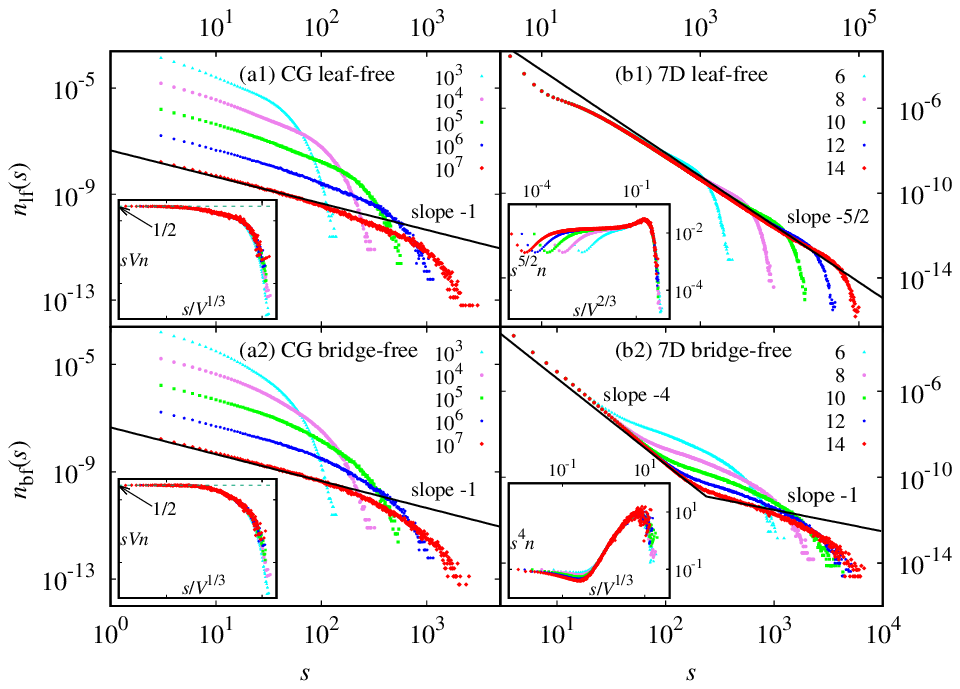}}

\title{Critical percolation clusters in seven dimensions and on a complete graph}
\date{\today}
\author{Wei Huang}
\affiliation{Hefei National Laboratory for Physical Sciences at Microscale and 
Department of Modern Physics, University of Science and Technology of China, 
Hefei, Anhui 230026, China}
\author{Pengcheng Hou}
\affiliation{Hefei National Laboratory for Physical Sciences at Microscale and 
Department of Modern Physics, University of Science and Technology of China, 
Hefei, Anhui 230026, China}
\author{Junfeng Wang}
\email{wangjf@hfut.edu.cn}
\affiliation{School of Electronic Science and Applied Physics, Hefei University 
of Technology, Hefei, Anhui 230009, China}
\author{Robert M. Ziff}
\email{rziff@umich.edu}
\affiliation{Department of Chemical Engineering, University of Michigan, Ann Arbor MI 48109-2136 USA}
\author{Youjin Deng}
\email{yjdeng@ustc.edu.cn}
\affiliation{Hefei National Laboratory for Physical Sciences at Microscale and 
Department of Modern Physics, University of Science and Technology of China, 
Hefei, Anhui 230026, China}
\affiliation{CAS Center for Excellence and Synergetic Innovation Center in Quantum Information and Quantum Physics, University of Science and Technology of China, Hefei, Anhui 230026, China}

\begin{abstract}
We study critical bond percolation on a seven-dimensional (7D) hypercubic lattice with periodic boundary conditions and on the complete graph (CG)
of finite volume (number of vertices) $V$.  We numerically confirm that for both cases, the critical number density $n(s,V)$  of clusters of size $s$ 
obeys a scaling form $n(s,V) \sim s^{-\tau} \tilde{n} (s/V^{d^*_{\rm f}})$ with identical volume fractal dimension $d^*_{\rm f}=2/3$ 
and exponent $\tau = 1+1/d^*_{\rm f}=5/2$.
We then classify occupied bonds into {\em bridge} bonds, which includes {\em branch} and {\em junction} bonds, and {\em non-bridge} bonds;
a bridge bond is a branch bond if and only if its deletion produces at least one tree.
Deleting branch bonds from percolation configurations produces {\em leaf-free} configurations, whereas,
deleting all bridge bonds leads to {\em bridge-free} configurations (blobs).
It is shown that the fraction of non-bridge (bi-connected) bonds vanishes $\rho_{\rm n, CG}$$\rightarrow$0 for large CGs, 
but converges to a finite value $ \rho_{\rm n, 7D} =0.006 \, 193 \, 1(7)$ for the 7D hypercube.
Further, we observe that while the bridge-free dimension $d^*_{\rm bf}=1/3$ holds for both the CG and 7D cases,
the volume fractal dimensions of the leaf-free clusters are different: 
$d^*_{\rm \ell f, 7D} = 0.669 (9) \approx 2/3$ and  $d^*_{\rm \ell f, CG} = 0. 333 7 (17) \approx 1/3$. 
We also study the behavior of the number and the size distribution of leaf-free and bridge-free clusters. 
For the number of clusters, we numerically find the number of leaf-free and bridge-free clusters on the CG scale as $\sim \ln V$,
while for 7D they scale as $\sim V$.
For the size distribution, we find the behavior on the CG is governed by a modified Fisher exponent $\tau^{\prime}=1$, 
while for leaf-free clusters in 7D it is governed by Fisher exponent $\tau=5/2$.  The size distribution of bridge-free clusters in 7D 
displays two-scaling behavior with exponents $\tau=4$ and $\tau^{\prime}=1$. 
Our work demonstrates that the geometric structure of high-dimensional percolation clusters cannot be fully accounted for 
by their complete-graph counterparts.

\end{abstract}
\maketitle

\section{Introduction}
\label{Introduction}
At the threshold $p_c$, the percolation process leads to random, scale-invariant geometries that have become
paradigmatic in theoretical physics and probability theory~\cite{StaufferAharony1994,Grimmett1999,BollobasRiordan2006,AraujoEtAl14}.
In two dimensions, a host of exact results are available.  
The bulk critical exponents $\beta = 5/36$ (for the order parameter) and $\nu=4/3$ (for the 
correlation length) are predicted by Coulomb-gas arguments~\cite{Nienhuis1987}, conformal 
field theory~\cite{Cardy1987}, and SLE theory~\cite{LawlerSchrammWerner2001},
and are rigorously confirmed in the specific case of triangular-lattice site 
percolation~\cite{SmirnovWerner2001}.
In high dimensions, above the upper critical dimensionality $d_{\rm u}=6$, the mean-field values $\beta=1$ and $\nu=1/2$ 
are believed to hold~\cite{Aharony1984,HaraSlade1990,Fitzner2017}. For dimensions $2 < d < d_{\rm u}$,  exact values of exponents are unavailable,
and the estimates of  $\beta$ and $\nu$ depend primarily upon numerical simulations~\cite{Wang13,Xu2014si,Paul2001}.

Two simple types of lattices or graphs have been used to model infinite-dimensional systems: 
the Bethe lattice (or Cayley tree), and the complete graph (CG). 
A Bethe lattice is a tree on which each site has a constant number $z$ of branches, and 
the percolation process becomes a branching process with threshold $p_c=1/(z-1)$~\cite{StaufferAharony1994}.
On a finite CG of $V$ sites, there exist $V(V-1)/2$ links between all pairs of sites;
the bond probability is denoted as $p$ with $p_c=1/(V-1)$~\cite{ErdosRenyi1960,Stepanov1970}.
In the thermodynamic limit of $V \rightarrow \infty$,  bond percolation on the Bethe lattice and the CG 
both lead to the critical exponents $\beta =1$ and $\nu=1/2$.  In this limit, the CG becomes essentially the  Bethe lattice because the probability of forming a loop vanishes.  
In this paper we use the CG to compare with 7D lattice percolation because the CG is isotropic, while the Bethe lattice has a very large surface of non-isotropic sites for finite systems.

In the Monte-Carlo study of critical phenomena, finite-size scaling (FSS) provides a key computational tool for estimating critical exponents.  
Consider bond percolation on a $d$-dimensional lattice with linear size $L$,
in which each link of a lattice is occupied with probability $p$.  FSS predicts that near the percolation threshold $p_c$, 
the largest-cluster size $C_1 $ scales  asymptotically as 
\begin{equation}
C_1 (t, L) = L^{d_{\rm f}} \tilde{C}_1 (tL^{y_t}) \hspace{5mm} ( t= p-p_c) \; ,
\label{eq:FSS_C1} 
\end{equation}
where the thermal and magnetic exponents, $y_t$ and $d_{\rm f}$, are related to the bulk exponents as $y_t=1/\nu$ and $d- d_{\rm f} =\beta/\nu$,
and $\tilde{C}_1$ is a universal function (if we include metric factors in its argument and coefficient).
Exponent $d_{\rm f}$ is the standard fractal dimension of the clusters.
 
Although well established for $d < d_{\rm u}$, FSS for $d >d_{\rm u}$ is surprisingly subtle and depends on boundary conditions~\cite{kennaBerche2016}. 
For instance, at the percolation threshold $p_c$, it is predicted that the fractal  dimension is 
$d_{\rm f}=4$~\cite{Hara2008} and $2d/3$~\cite{HeydenreichHofstad2007} for systems with free and periodic boundary conditions, respectively.
At $p_c$, the largest cluster size on the CG scales as $C_1 \propto V^{2/3}$, implying a volume fractal dimension $d^*_{\rm f}=2/3$~\cite{Janson1993}.
An interesting question arises: how well does the CG describe 
other aspects of high-dimensional percolation $d> d_{\rm u}$?

In this work, we simulate bond percolation on the 7D hypercubic lattice with periodic boundary conditions, and on the CG.  
We numerically confirm the FSS of the size of the largest critical cluster $C_1 \propto V^{2/3}$ for both systems.
Furthermore, we show that the cluster number $n(s,V)$ of size $s$ per site at the critical point obeys a universal scaling form 
\begin{equation}
n(s,V) = n_0 s^{-\tau} \tilde{n} (s/V^{d^*_{\rm f}}) \; ,
\label{eq:FSS_ns}
\end{equation}
where  $d^*_{\rm f}=2/3$ is a volume fractal dimension, equal to $d_{\rm f}/d$ for spatial systems, exponent $\tau = 1+1/d^*_{\rm f} =5/2$, 
$n_0$ is a non-universal constant, and $\tilde{n}$ is a universal function (if we include a metric factor in its argument). We numerically confirm that
$n_0$ for the CG is equal to $(2\pi)^{-1/2} \approx 0.3989$~\cite{Krapivsky05}, while for 7D $n_0 = 0.527(7)$ which is definitely higher than $(2\pi)^{-1/2}$.

\begin{figure}
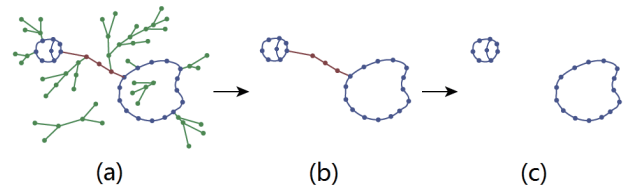

 \cartoonfFig
  \caption{Sketch of the types of bonds and clusters considered here:  (a) the initial complete clusters, (b) leaf-free clusters in which all trees containing branch bonds (green) have been removed, including the removal of entire clusters if they are completely trees, and (c) bridge-free clusters or blobs (blue), in which all junction bonds (brown) as well as branch bonds have been removed.  
 }
  \label{Fig:cartoonfFig}
\end{figure}

We then consider a natural classification of the occupied bonds of a percolation configuration 
and study the FSS of the resulting clusters, following the procedure in Ref.~\cite{XuWangZhouTimDeng2014}.
The occupied bonds are divided into {\em bridge} bonds and {\em non-bridge} bonds, 
and bridge bonds are further classified as {\em branch} bonds and {\em junction} bonds.
A bridge bond is an occupied bond whose deletion would break a cluster into two.
The bridge bond is a junction bond if neither of the two resulting clusters is a tree; otherwise, it is a branch bond.
Deleting all branches from percolation configurations leads to {\em leaf-free} configurations, which have no trees,
and further deleting junctions produces {\em bridge-free} configurations.  This process is shown schematically in Fig.\ \ref{Fig:cartoonfFig}. Other terminology is to call leaves and trees ``dangling ends," and the bridge-free clusters ``bi-connected" or  ``blobs."  The junction bonds are called ``red bonds" when they are along the conduction path of the system, and in general not all junction bonds are red bonds.

We find that while the fraction of non-bridge bonds vanishes $\rho_{\rm n, CG} (V \rightarrow \infty) =0$ for 
infinitely large CGs at criticality, it converges to a finite thermodynamic value $ \rho_{\rm n, 7D} =0.006 \, 193 \, 1(7)$ for $7D$ percolation.
This implies that in contrast to the complete-graph case, the number of loops or blobs in finite-$d$ critical percolation configurations 
is proportional to volume $V = L^d$. We further determine the volume fractal dimensions of the leaf-free and bridge-free clusters as 
$d^*_{\rm \ell f, 7D} = 0.669(9) \approx 2/3$ and $d^*_{\rm bf, 7D} = 0.332(7) \approx 1/3$  for 7D, 
and $d^*_{\rm \ell f, CG} = 0.3337(17) \approx 1/3$ and $d^*_{\rm bf, CG} = 0.3337(15) \approx 1/3$ on the CG.
While the bridge-free clusters apparently share the same fractal dimension $d^*_{\rm bf}=1/3$ for the two systems, 
the leaf-free clusters have dramatically different fractal dimensions and scaling exponents.

Moreover, we confirm that the number of leaf-free and bridge-free clusters on the CG are proportional to $\ln V$ on the CG,  
while we find they are proportional to $V$ in 7D. Further, we find the behavior of the size distribution of leaf-free and bridge-free clusters on the CG are governed by 
a modified Fisher exponent $\tau^{\prime}=1$ which is related to the fact that number of clusters is proportional to $\ln V$, 
while the distribution for leaf-free clusters in 7D is governed by Fisher exponent $\tau=5/2$. However, the size distribution of bridge-free clusters in 7D 
displays two-scaling behavior with exponents $\tau=4$ and $\tau^{\prime}=1$ respectively. 

The remainder of this paper is organized as follows.
Section~\ref{Simulation} describes the simulation algorithm and sampled quantities.
In Sec.~\ref{Results}, 
the Monte-Carlo data are analyzed, and results for bond densities, various fractal dimensions, number of clusters as well 
as the size distribution $n(s,V)$ are presented.
A discussion of these results is given in Sec.~\ref{Discussion}.

\section{Simulation}
\label{Simulation}
\subsection{Model}
\label{Model}
We study critical bond percolation on the 7D hypercubic 
lattice with periodic boundary conditions and on the CG, 
at their thresholds $p_c=0.0786752(3)$~\cite{Grassberger03} and 
$p_c=1/(V-1)$, respectively. 
At the first step of the simulation, 
we produce the configurations of the complete clusters.
On the basis of these clusters, all occupied bonds are classified into three types: branch,
junction, and non-bridge. 
Definitions of these terminologies have been given 
in~\cite{XuWangZhouTimDeng2014} for two-dimensional percolation,  
and can be transplanted intactly to the present models. A {\em leaf} is defined as a site which is adjacent to precisely one occupied bond,
and a `leaf-free' configuration is then defined as a configuration 
without any leaves.
In actual implementation, we generate the 
{\em leaf-free} configuration via the following iterative 
procedure, often referred to as {\em burning}. 
For each leaf, we delete its adjacent bond. If this procedure generates 
new leaves, we repeat it until no leaves remain. 
The bonds which are deleted during this iterative process 
are precisely the branch bonds, and the remaining bridge bonds in the leaf-free configurations 
are the junction bonds. 
If we further delete all junction bonds from the leaf-free configurations,
then we obtain the bridge-free configurations.
We note that the procedure of identifying the non-bridge bonds 
from a leaf-free configuration can be time consuming, since it involves 
the check of global connectivity.
In Sec.~\ref{Algorithm}, 
we describe the algorithm we used to carry this out efficiently.

We simulated multiple system sizes for each model.
For 7D lattice percolation, 
we simulated the linear system sizes $L=5$, 6, 7, 8, 9, 10, 11, 12,
with no less than $10^6$ independent samples for each $L$.
For the CG, we simulated volumes $V=2^{8}$, $2^{9}$, $2^{10}$,..., $2^{20}$ number of sites, generating at least $10^7$ independent samples for each $V$.

\subsection{Algorithm}
\label{Algorithm}

Unlike in the planar case~\cite{XuWangZhouTimDeng2014}, we cannot take advantage of the associated loop configurations, so the algorithm for 2D is not suitable for percolation clusters in higher dimensions or on CGs. 
To identify non-bridge bonds within leaf-free clusters, we developed an algorithm based upon a breadth-first growth 
algorithm, which could be seen as a special case of the matching algorithm~\cite{Moukarzel96,Moukarzel98}. 
Different from~Ref.\cite{Moukarzel98} in which loops between two points far apart have to be identified dynamically, 
we just need to identify all the loops within a cluster, which results in a simpler version of the algorithm.

Consider an arbitrary graph $G=(N,E)$ of $N$ sites (vertices) labeled as $i = 1,...,N$ connected by a set of $|E|$ edges. 
Similarly to the matching algorithm, we implement a directed graph $G^*$ as an auxiliary representation of the system to identify loops within a leaf-free cluster. 
To represent the directed graph $G^*$, we set an array called ${\mathcal D}$ of size $N$; if a site $j$ points to another site $i$, then we assign 
${\mathcal D}(j)=i$. 

Starting from a leaf-free cluster, we perform breadth-first search from site $i_0$, and assign ${\mathcal D}(i_0)=i_0$. 
We add one edge at one step of the search to the site $i_1$ and assign ${\mathcal D}(i_1)=i_0$.  Before adding a new edge to graph $G^*$, we 
check whether the new site $i_n$ on the growth process has been visited before; if it has, 
then the new edge will close a loop. Once a loop forms at $i_n$, we follow the arrows backwards until the two backtracking paths meet.
In this way all the edges of the loop are identified. After identifying all the non-bridge bonds on the loop, 
we assign the value of the starting point of the loop to all the elements of $G^*$ that belong to the loop. 
Once all the edges in a loop have been identified, we continue to perform breadth-first growth on the leaf-free cluster and identify all the remaining non-bridge bonds. 

For critical percolation on CGs, the percolation threshold $p_c$ equals to $1/(V-1)$, and thus $p_c$ becomes small
as $V$ becomes large. Therefore, at the critical probability, most of edges are unoccupied and storing the occupied edges instead of all the edges in the system saves a large amount of computer memory.

The small $p_c$ for both 7D and CG implies that many random numbers and many operations would be needed if all the potentially occupied 
edges were visited to decide whether they are occupied or not. The simulation efficiency would drop quickly as the coordination 
number increases. In our algorithm we follow a more efficient procedure~\cite{Deng02,Deng05}. We define $P(i)\equiv(1-p_c)^{i-1}p_c$ to be 
the probability that the first $(i-1)$ edges are empty (unoccupied) while the $i$th edge is occupied. The cumulative probability distribution 
$F(i)$ is then
\begin{equation}
  F(i) = \sum_{i^{\prime}=1}^eP(i^{\prime}) = 1-(1-p_c)^i.
  \label{Eq:distribution}
  \end{equation}
which gives the probability that the number of bonds to the first occupied edge is less than or equal to $i$.

Now, suppose the current occupied edge is the $i_0$th edge, one can obtain the next-to-be occupied edge as the $(i_0+i)$th edge by drawing a
uniformly distributed random number $0\leq r <1$ and determining the value of $i$ such that
\begin{equation}
  F(i-1) \leq r < F(i)
  \label{Eq:position}
\end{equation}
Solving equation (\ref{Eq:position}) we get
\begin{equation}
  i = 1 + \lfloor \ln(r)/\ln(1-p_c) \rfloor \; .
  \label{Eq:sovle}
\end{equation}
Thus, instead of visiting all the potentially occupied edges, we directly jump to the next-to-be occupied edge in the system, skipping  all the unoccupied ones sequentially. This process is repeated 
until the state of all edges in the system have been decided. By this method, the simulation efficiency is significantly improved for small $p_c$. This procedure is especially beneficial for the CG, where the total number of bonds  $V(V-1)/2$ can be huge.

\subsection{Measured quantities}
\label{Measured quantities}
We measured the following observables in our simulations:
\begin{enumerate}

\item The mean branch-bond density $\rho_{\rm b} \coloneqq N_{\rm b}/(|E|p_c)$ where $N_{\rm b}$ is the number of branch bonds and $|E|$ the total number of edges.
         Analogously, the mean junction-bond density $\rho_{\rm j}$ 
         and the mean non-bridge-bond density $\rho_{\rm n}$. It is clear that $\rho_{\rm b}+\rho_{\rm j}+\rho_{\rm n}=1$.
\item The mean number $N$ of complete clusters,  $N_{\rm {\ell f}}$ of the leaf-free clusters, 
         and $N_{\rm {bf}}$ of the bridge-free clusters.
         Note that while an isolated site is counted as a complete cluster, it is burned out by definition 
         and is not counted as a leaf-free or bridge-free cluster.
\item The number density $n(s)$ of clusters of size $s$ for the complete clusters, $n_{\rm {\ell f}}(s)$ for the leaf-free clusters, and 
         $n_{\rm {bf}}(s)$ for the bridge-free clusters.
\item The mean size $C$ of the largest complete cluster, $C_{\rm {\ell f}}$ of the largest leaf-free cluster, and $C_{\rm {bf}}$ of the largest bridge-free cluster.
\end{enumerate}

\section{Results}
\label{Results} 

\subsection{Bond densities}
\label{Bond densities} 

In the definition of bond densities $\rho_i$, the number of bonds is not only divided by $|E|$, but also by $p_c$, so they represent the fraction of each kind of bond, 
and $\rhobranch+\rhojunction+\rhononbridge = 1$. We fit our Monte-Carlo data for the densities $\rhojunction$, $\rhobranch$ and $\rhononbridge$ to the finite-size scaling ansatz
\begin{equation}
  \rho = \rho_0 + V^{-y}(a_0 + a_1 V^{-y_1} + a_2 V^{-y_2})
  \label{Eq:Bond_Density}
\end{equation} 
where $\rho_0$ is the critical value of bond density in the thermodynamic limit, $y$ is the leading correction exponent and $y_j$ ($j=1,2$) are sub-leading exponents. 

As a precaution against correction-to-scaling terms that we fail to include in the fitting ansatz, we impose a lower cutoff $L>L_{\rm min}$ on the data points admitted 
in the fit, and we systematically study the effect on the $\chi^2$ value of increasing $L_{\rm min}$. Generally, the preferred fit for any given ansatz corresponds 
to the smallest $L_{\min}$ for which the goodness of fit is reasonable and for which subsequent increases in $L_{\min}$ do not cause the $\chi^2$ value to drop by vastly 
more than one unit per degree of freedom. In practice, by ``reasonable'' we mean that $\chi^2/\mathrm{DF}\lessapprox 1$, where DF is the number of degrees of freedom.

\begin{table}
\begin{center}
  \begin{tabular}[t]{|l|l|l|l|l|l|l|}
     \hline
        &     $\rho$      & $\rho_{0}$        & $y$        &  $a_0$      & {\small $V_{\rm min}$}    & {\small DF/$\chi^2$} \\
     \hline
       {\multirow{9}{*}{CG}} & {\multirow{3}{*}{$\rhobranch$}}     &~1.000\,000(1)     &0.667(2)       &-2.01(4)          &$2^9$       &~~7/5 \\
     \cline{3-7}
     &                                                             &~0.999\,999(1)     &0.669(3)       &-2.06(7)          &$2^{10}$    &~~6/4  \\     
     \cline{3-7}
     &                                                             &~0.999\,999(2)     &0.669(4)       &-2.05(12)         &$2^{11}$    &~~5/4 \\  
     \cline{2-7}
     &                        {\multirow{3}{*}{$\rhojunction$}}    &-0.000\,000\,1(1)  &0.661(1)       &~0.151(2)         &$2^9$       &~~7/9 \\
     \cline{3-7}
     &                                                             &-0.000\,000\,1(1)  &0.662(2)       &~0.154(4)         &$2^{10}$    &~~6/8 \\
     \cline{3-7}
     &                                                             &-0.000\,000\,1(1)  &0.660(3)       &~0.150(7)         &$2^{11}$    &~~5/7 \\
     \cline{2-7}
     &                         {\multirow{3}{*}{$\rhononbridge$}}  &~0.000\,000\,0(2)  &0.666\,8(5)    &~1.83(1)          &$2^9$       &~~7/11 \\
     \cline{3-7}
     &                                                             &~0.000\,000\,0(3)  &0.666\,6(8)    &~1.83(2)          &$2^{10}$    &~~6/10 \\
     \cline{3-7}
     &                                                             &~0.000\,000\,4(4)  &0.665\,1(12)   &~1.79(3)          &$2^{11}$    &~~5/8 \\
     \hline 
       {\multirow{6}{*}{7D}} & {\multirow{2}{*}{$\rhobranch$}}     &~0.985\,330\,2(2)  &$2/3$          &-0.017(12)        &$7^7$       &~~3/5 \\
     \cline{3-7}
     &                                                             &~0.985\,330\,5(3)  &$2/3$          &-0.038(23)        &$8^7$       &~~2/4 \\     
     \cline{2-7}
                              & {\multirow{2}{*}{$\rhojunction$}}  &~0.008\,476\,4(2)  &0.661(2)       &~1.53(7)          &$6^7$       &~~3/2 \\
     \cline{3-7}
     &                                                             &~0.008\,476\,5(3)  &0.664(8)       &~1.60(19)         &$7^7$       &~~2/2 \\     
     \cline{2-7}
                              & {\multirow{2}{*}{$\rhononbridge$}} &~0.006\,193\,1(3)  &0.671(4)       &-1.77(10)         &$6^7$       &~~3/1 \\
     \cline{3-7}
     &                                                             &~0.006\,193\,1(5)  &0.671(11)      &-1.75(29)         &$7^7$       &~~2/1\\     
     \hline    

  \end{tabular}
  \caption{Fit results for branch-bond density $\rhobranch$, junction-bond density $\rhojunction$, 
  and non-bridge-bond density $\rhononbridge$ on the CG and in 7D.}
  \label{Tab:Bond_Density}
 \end{center}
 \end{table}

\begin{figure}
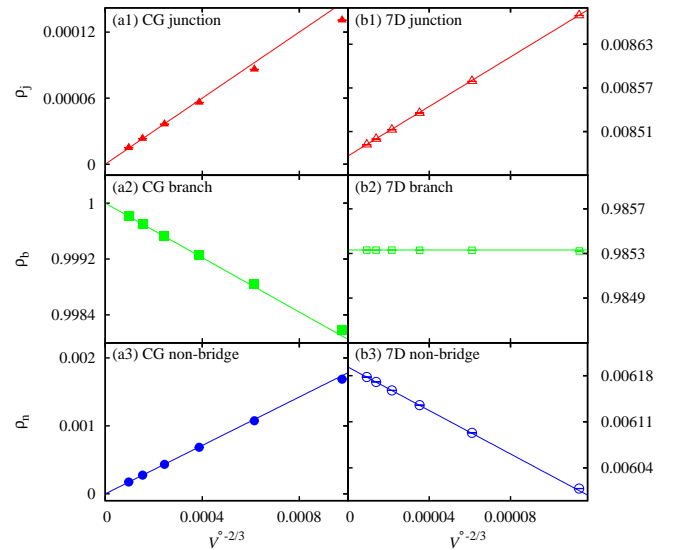

  \bondensiFig
  \caption{Junction-bond densities $\rhojunction$ (top), branch-bond density $\rhobranch$ (middle), 
  and non-bridge-bond density $\rhononbridge$ (bottom) vs.\ $V^{-2/3}$ on the CG (left) and in 7D (right).
    For the CG, from top to bottom, the intercept respectively corresponds to values $0, 1, 0$.
    For 7D, from top to bottom, the three intercepts respectively correspond to values $0.008\,476\,5$, $0.985\,330\,4$, and $0.006\,193\,1$.
   The statistical error of each data point is smaller than the symbol size. The straight lines are drawn simply to guide the eye.
 }
  \label{Fig:bondensiFig}
\end{figure}

In the fits for the CG data, we tried various values for $y_1$ and $y_2$ terms, and found fixing $y_1=1/3$ and $y_2 = 2/3$ lead to the most stable fitting results. 
Leaving $y$ free in the fits of $\rhobranch$, $\rhojunction$ and $\rhononbridge$ on the CG, we estimate $y= 0.669(6), 
0.661(6), 0.666(3)$ respectively, which are all consistent with $2/3$.	 

From the fits, we estimate for the CG that $\rhobranchLim=0.999\,999(3) \approx 1$, $\rhojunctionLim=-0.000\,000\,1(2) \approx 0$ and $\rhononbridgeLim=0.000\,000\,3(7) \approx 0 $. We note that 
$\rhobranchLim+\rhojunctionLim+\rhononbridgeLim=1$ within error bars, as expected. As the system tends to infinity, we conclude that $\rhobranchLim$ equals 
 1 while $\rhojunctionLim$ and $\rhononbridgeLim$ are equal to 0 which agrees with the findings in Ref.~\cite{ErdosRenyi1960}. This conclusion is consistent with 
the scenario for percolation on the Bethe lattice (Cayley tree) where all of the bonds are branches. 
Besides, we estimate $a_0$ for $\rhobranch$, $\rhojunction$ and $\rhononbridge$ on the CG as  $-2.05(18), 
0.150(11), 1.80(7)$ respectively. We note that the sum of these values is equal to 0 within error bars, as we expect. 

For bond densities in 7D, we fixed $a_2 = 0$ and tried various values for $y_1$, and finally fixed $y_1 = 1$. Leaving $y$ free in the fits of $\rhojunction$ and $\rhononbridge$, 
we obtain $y= 0.664(12)$ and $0.671(17) $ respectively which are consistent with $2/3$. On this basis, we conjecture that the leading finite-size correction exponent 
to bond densities in both models are $2/3$. For $\rhobranch$, by contrast, we were unable to obtain stable fits with $y$ free. 
Fixing $y=2/3$, the resulting fits produce estimates of $a$ that are consistent with zero. 
In fact, we find $\rhobranch$ is consistent with 0.985\,330 for all $V\geq 10^7$ ($L\geq 10$). 

From the fits, we estimate densities of  $\rhobranchLim = 0.985\,330\,4(6)$, $\rhojunctionLim = 0.008\,476\,5(5)$ and $\rhononbridgeLim = 0.006\,193\,1(7)$. We note that 
$\rhobranchLim+\rhojunctionLim+\rhononbridgeLim = 1$ within error bars. We also note that the estimates of $a_0$ for $\rhojunction$ and $\rhononbridge$ 
are equal in magnitude and opposite in sign, which is as expected given that $a_0$ for $\rhobranch$ is consistent with zero. The fit details are summarized in 
Table~\ref{Tab:Bond_Density}. 

In Fig.~\ref{Fig:bondensiFig}, we plot $\rhobranch$, $\rhojunction$ and $\rhononbridge$ of CG and 7D vs.\ $V^{-2/3}$. For the CG, the plot clearly demonstrates that the 
leading finite-size corrections for $\rhobranch$, $\rhojunction$ and $\rhononbridge$ are governed by exponent $-2/3$, and $\rhobranch$ tends to 1 while 
$\rhojunction$ $\rhononbridge$ tend to 0 when the system tends to infinity. For 7D, the plot clearly demonstrates that the leading finite-size 
corrections for $\rhojunction$ and $\rhononbridge$ are governed by exponent $-2/3$, while essentially no finite-size dependence can be discerned for $\rhobranch$.

\subsection{Fractal dimensions of clusters}
\label{Fractal dimensions of clusters}

In this subsection, we estimate the volume fractal dimensions $d^*_{\rm f}$, $d^*_{\ell {\rm f}}$, and $d^*_{\rm bf}$ from 
the observables $\percolationCluster$, $\leafFreeCluster$ and $\bridgeFreeCluster$, respectively,
which are fitted to the finite-size scaling ansatz
\begin{equation}
  \scrO = c_0 + V^{d^*_\scrO}(a_0 + a_1 V^{-y_1} + a_2 V^{-y_2} + a_3 V^{-y_3})
  \label{Eq:Fractal_dimensions}
\end{equation}
where $d^*_\scrO$ denotes the appropriate volume fractal dimension.
The fit results are reported in Table~\ref{Tab:FD_Clusters}. In the reported fits we set $c_0 = 0$ identically, since leaving it free produced 
estimates for it consistent with zero. 

\begin{figure}
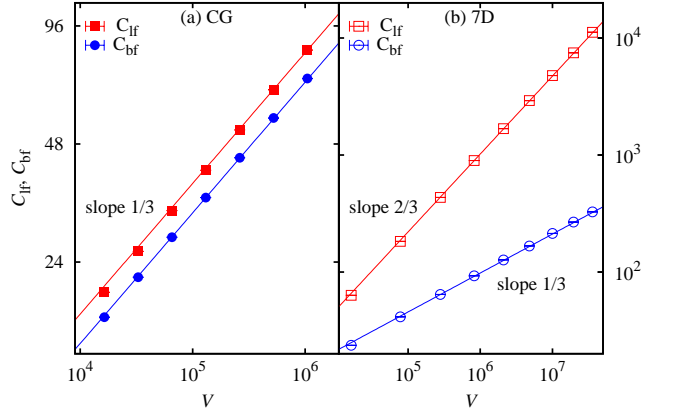

  \fractalfFig
  \caption{(a) Log-log plot of $\leafFreeCluster$, $\bridgeFreeCluster$ vs.\ $V$ on the CG. The two solid lines have slopes 1/3.
           The statistical error of each data point is smaller than the symbol size. The straight lines are simply to guide the eye.
           (b) Log-log plot of $\leafFreeCluster$, $\bridgeFreeCluster$ vs.\ $V$ in 7D. The two solid lines have slopes 2/3 and 1/3 respectively.  
           The statistical error of each data point is smaller than the symbol size. The straight lines are simply to guide the eye.}
  \label{Fig:fractalfFig}
\end{figure}

For percolation on the CG, we fix $y_1=1/3$, $y_2=2/3$ and $y_3=1$. We estimate $d^*_{\rm f}$=0.666\,4(5), which is consistent with the predicted volume fractal dimension 
of percolation clusters, $d^*_{\rm f}= 2/3$. Besides, we estimate $d^*_{\ell {\rm f}}=0.333\,7(17)$ and $d^*_{\rm bf}=0.333\,7(15)$. In Fig.~\ref{Fig:fractalfFig}(a), 
we plot $\leafFreeCluster$ and $\bridgeFreeCluster$ on the CG vs.\ $V$. 
We conjecture both volume fractal dimensions of $\leafFreeCluster$ and $\bridgeFreeCluster$ on the CG are equal to $1/3$. 

For percolation in 7D, we fix $y_1=1/3$, $y_2=2/3$ and $a_3=0$. We estimate $d_{\rm f}$=0.669(9), which is consistent with the volume fractal dimension of percolation 
clusters, $d^*_{\rm f}= 2/3$. Furthermore, we estimate $d^*_{\ell {\rm f}}$=0.669(9) and $d^*_{\rm bf}$=0.332(7). In Fig.~\ref{Fig:fractalfFig}(b), we plot $\leafFreeCluster$ and 
$\bridgeFreeCluster$ of 7D vs.\ $V$ to illustrate our estimates for $d^*_{\ell {\rm f} }$ and $d^*_{\rm bf}$. We conjecture that $d^*_{\ell {\rm f} }=2/3$ 
while $d^*_{\rm bf}=1/3$.

As our numerical analysis shows, $d^*_{\rm f}$ and $d^*_{\rm bf}$  are consistent on the CG and in 7D,  
while $d^*_{\ell {\rm f}}$ is not consistent for the two systems.

\begin{table}[htbp]
  \begin{tabular}[t]{|l|l|l|l|l|l|l|}
     \hline
             &     $\scrO$      &  $d^*_\scrO$        & $a_0$        &  $a_1$       & $V_{\rm min}$ & DF/$\chi^2$ \\
     \hline
       {\multirow{9}{*}{CG}} & {\multirow{3}{*}{$\percolationCluster$}}     &0.666\,5(1) &0.942(2)   & -0.21(2)     &$2^8$       &~~8/5 \\
     \cline{3-7}
     &                                                                      &0.666\,4(2) &0.943(3)   & -0.22(4)     &$2^9$       &~~7/5 \\     
     \cline{3-7}
     &                                                                      &0.666\,4(3) &0.944(4)   & -0.24(7)     &$2^{10}$    &~~6/5 \\  
     \cline{2-7}
     &                        {\multirow{3}{*}{$\leafFreeCluster$}}         &0.333\,1(3) &0.834(4)   & -1.22(5)     &$2^8$       &~~8/14 \\
     \cline{3-7}
     &                                                                      &0.333\,5(5) &0.830(6)   & -1.14(8)     &$2^9$       &~~7/12 \\
     \cline{3-7}
     &                                                                      &0.333\,8(7) &0.826(9)   & -1.07(14)    &$2^{10}$    &~~6/11 \\
     \cline{2-7}
     &                         {\multirow{3}{*}{$\bridgeFreeCluster$}}      &0.333\,2(3) &0.700(3)   & -0.49(4)     &$2^8$       &~~8/12 \\
     \cline{3-7}
     &                                                                      &0.333\,5(5) &0.697(5)   & -0.44(7)     &$2^9$       &~~7/11 \\
     \cline{3-7}
     &                                                                      &0.333\,8(7) &0.693(7)   & -0.38(12)    &$2^{10}$    &~~6/10 \\
     \hline 
       {\multirow{6}{*}{7D}} & {\multirow{2}{*}{$\percolationCluster$}}     &0.665(2)    &1.17(5)    &~~-           &$5^7$       &~~3/4 \\
     \cline{3-7}
     &                                                                      &0.669(6)    &1.08(13)   &~~-           &$6^7$       &~~2/3 \\  
     \cline{2-7}
                              & {\multirow{2}{*}{$\leafFreeCluster$}}       &0.665(2)    &0.107(4)   &~~-           &$5^7$       &~~3/4 \\
     \cline{3-7}
     &                                                                      &0.669(6)    &0.098(12)  &~~-           &$6^7$       &~~2/3 \\ 
     \cline{2-7}
                              & {\multirow{2}{*}{$\bridgeFreeCluster$}}     &0.332(2)    &1.01(3)    &~~-           &$5^7$       &~~3/4 \\
     \cline{3-7}
     &                                                                      &0.332(5)    &1.01(9)    &~~-           &$6^7$       &~~2/4 \\     
     \hline      

  \end{tabular}
  \caption{Fit results for the size of the largest complete cluster $\percolationCluster$, 
  the size of the largest leaf-free cluster $\leafFreeCluster$, and the size of the largest bridge-free cluster $\bridgeFreeCluster$.}
  \label{Tab:FD_Clusters}
\end{table} 

\subsection{Number of clusters}
\label{Number of clusters}

\begin{figure}
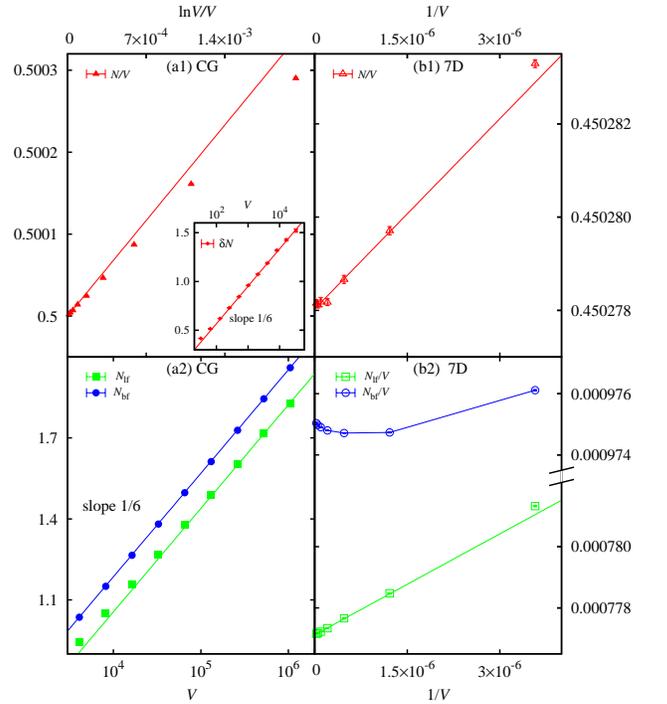

  \clusternFig
  \caption{ (a1) Plot of number density $N/V$ of complete clusters versus $(\ln V)/V$ on the CG. The right bottom inset is the semi-log plot of $\delta N$ versus $V$. 
            (a2) Semi-log plot of total number $N_{\rm \ell f}$, $N_{\rm bf}$ of leaf-free and bridge-free clusters versus $V$ on the CG.
            (b1) Plot of the number density $N/V$ of complete clusters versus $1/V$ in 7D.
            (b2) Plot of number density $N_{\rm \ell f}/V$, $N_{\rm bf}/V$ of leaf-free and bridge-free clusters versus $1/V$ in 7D.}
  \label{Fig:clusternFig}
\end{figure}

According to~\cite{ErdosRenyi1960}, the average number of complete clusters at criticality on the CG satisfies
\begin{equation}
  N = V/2 + {\rm O}(\ln V)
  \label{Eq:Number_Cluster1}
\end{equation}
To verify this theorem, we study the number of complete clusters at criticality on the CG and fit the data to the ansatz
\begin{equation}
  N =  a_0V + a_1 \ln V + a_2 + a_3 V^{-1}
  \label{Eq:Number_Cluster2}
\end{equation}
The resulting fits are summarized in Table~\ref{Tab:Nu_Clusters}. Leaving $a_3$ free, we find that $a_3$ is consistent with zero, suggesting that the last sub-leading correction
exponent in the ansatz might be even smaller than $-1$. We estimate $a_0 = 0.500\,000\,1(3) \approx 1/2 $ and $a_1 = 0.167(9)$.
In Fig.~\ref{Fig:clusternFig}(a1) we plot $ \delta N \coloneqq N-V/2$ vs.\ $V$ to illustrate the logarithmic correction of cluster number of critical percolation on the CG.  

We also study the number of leaf-free clusters and bridge-free clusters on the CG. 
Since the fraction of the junction bonds and the non-bridge bonds, $\rho_{\rm j}$ and $\rho_{\rm n}$, vanish as  ${\rm O}(V^{-2/3})$, 
the number of isolated sites after burning is approaching to $V$ with a correction term $V^{-2/3}$. 
Note that in our definitions of leaf-free and bridge-free clusters, we do not include these isolated sites. 
As a result, the $V$ term does not exist in $N_{\rm \ell f}$ or $N_{\rm b f}$.
We find that $N_{\rm \ell f}$ and $N_{\rm b f}$ grow slowly as $V$ tends to infinity. As suggested by~\cite{Luczak94},
we fit the data to the ansatz
\begin{equation}
 N_{\scrO} = a_0 + a_1 \ln V + a_2V^{-1/3}+a_3V^{-2/3}+a_4V^{-1} 
  \label{Eq:Number_Cluster3}
\end{equation} 
where $N_\scrO$ represents $N_{\rm \ell f}$ or $N_{\rm b f}$. 
We estimate $a_0 = -497(22), -0.358(22)$ and $a_1 = 0.1672(9), 0.1673(10)$ for $N_{\rm \ell f}$ and $N_{\rm b f}$ respectively, 
which means that both the coefficient of the logarithmic term of the number $N$ of leaf-free and of bridge-free clusters are consistent with the value $1/6$. 
Combined with the results for $N$, this suggests that the logarithmic term in (\ref{Eq:Number_Cluster1}) comes from clusters containing cycles.
We mention that in Ref.~\cite{Krapivsky05}, they find that average number of unicyclic components grows logarithmically with the system size as $\sim (1/6)\ln V$, which indicates
that most leaf-free and bridge-free clusters are unicyclic components.  
In Fig.~\ref{Fig:clusternFig}(a2), we plot $N_{\rm \ell f}$ and $N_{\rm b f}$ vs.\ $V$ to illustrate the logarithmic growth of number of leaf-free and bridge-free clusters.

\begin{table}[htbp]
  \begin{tabular}[t]{|l|l|l|l|l|l|}
     \hline
             &     $ $       & $a_0$        &  $a_1$       & $V_{\rm min}$ & DF/$\chi^2$ \\
     \hline
       {\multirow{9}{*}{CG}} & {\multirow{3}{*}{$N$}}                       &~0.500\,000\,1(1)      & 0.167(2)     &$2^8$       &~14/9 \\
     \cline{3-6}
     &                                                                      &~0.500\,000\,1(1)      & 0.166(4)     &$2^9$       &~13/8 \\     
     \cline{3-6}
     &                                                                      &~0.500\,000\,1(2)      & 0.167(6)     &$2^{10}$    &~12/7 \\  
     \cline{2-6}
     &                        {\multirow{3}{*}{$N_{\rm \ell f}$}}           &-0.487(4)              & 0.1665(2)    &$2^8$       &~8/14 \\
     \cline{3-6}
     &                                                                      &-0.491(5)              & 0.1667(4)    &$2^9$       &~7/13 \\
     \cline{3-6}
     &                                                                      &-0.503(9)              & 0.1675(6)    &$2^{10}$    &~6/10 \\
     \cline{2-6}
     &                         {\multirow{3}{*}{$N_{\rm bf}$}}              &-0.347(4)              & 0.1666(3)    &$2^8$       &~8/12 \\
     \cline{3-6}
     &                                                                      &-0.351(6)              & 0.1668(4)    &$2^9$       &~7/11 \\
     \cline{3-6}
     &                                                                      &-0.364(9)              & 0.1676(6)    &$2^{10}$    &~6/9 \\
     \hline 
       {\multirow{6}{*}{7D}} & {\multirow{2}{*}{$N$}}                       &0.450\,278\,03(4)      & 1.36(8)      &$7^7$       &~4/5 \\
     \cline{3-6}
     &                                                                      &0.450\,278\,06(5)      & 1.13(22)     &$8^7$       &~3/4 \\  
     \cline{2-6}
     &                        {\multirow{2}{*}{$N_{\rm \ell f}$}}           &0.000\,777\,127(4)     & 1.01(4)      &$9^7$       &~2/1 \\
     \cline{3-6}
     &                                                                      &0.000\,777\,130(6)     & 0.93(11)     &${10}^7$    &~1/1 \\
     \cline{2-6}
     &                        {\multirow{2}{*}{$N_{\rm bf}$}}               &0.000\,975\,129(8)     & -0.89(6)     &$8^7$       &~2/3 \\
     \cline{3-6}
     &                                                                      &0.000\,975\,142(15)    & -1.08(19)    &$9^7$       &~1/1 \\ 
     \hline          

  \end{tabular}
  \caption{Fit results for the number of complete clusters $N$, 
  the number of leaf-free clusters $N_{\rm \ell f}$, and the number of bridge-free clusters 
  $N_{\rm bf}$ on the CG and in 7D.}
  \label{Tab:Nu_Clusters}
\end{table} 

For the 7D case, the behavior of cluster numbers is much different from the CG. We find that cluster number densities of complete, leaf-free and bridge-free clusters tend to a finite limit. 
This is demonstrated in Fig.~\ref{Fig:clusternFig}(b1) and Fig.~\ref{Fig:clusternFig}(b2). We find that the excess cluster number~\cite{Ziff97} could be 
found for complete clusters and leaf-free clusters. For bridge-free clusters, however, the behavior is not linear which implies that the excess cluster concept does not apply here. 
To estimate the excess cluster number of complete clusters and leaf-free clusters, we fit the cluster number $N$ and $N_{\rm \ell f}$ of complete clusters and leaf-free clusters to the ansatz
\begin{equation}
  N_\scrO =  a_0 V + a_1
  \label{Eq:Excess_number}
\end{equation} 
where $N_\scrO$ represents $N$ or $N_{\rm \ell f}$. 
We find that $N$ and $N_{\rm \ell f}$ can be well fitted to the ansatz (\ref{Eq:Excess_number}), and estimate the 
density of complete clusters $a_0 = 0.450\,278\,06(9)$, the excess cluster number of complete clusters $a_1 = 1.18(39)$, the density of leaf-free clusters
$a_0 = 0.000\,777\,130(13)$, and the excess cluster number of leaf-free clusters $a_1 = 0.93(18)$. 
As illustrated by Fig.~\ref{Fig:clusternFig}(b2),  the number density $N_{\rm bf}/V$ of bridge-free clusters does not scale monotonically as $V$ increases. 
It will be shown later  that for large size $s$, the cluster-size distribution $n_{\rm bf} (s,V)$  in 7D displays similar behavior as that 
on CG; see Figs.~\ref{Fig:distribfFig} (a2)-(b2).
Motivated by this observation, we conjecture that in addition to a term $\sim V$, the cluster number $N_{\rm bf}$ has a logarithmic term $\sim \ln V$.
On this basis, we fit the $N_{\rm bf}$  data by the ansatz
\begin{equation}
  N_{\rm bf} =  a_0 V + a_1 \ln V + a_2 \; ,
  \label{Eq:bf_density}
\end{equation}
and obtain the density of bridge-free clusters $a_0 = 0.000\,975\,139(27)$, $a_1=-0.89(10)$ and $a_2=12(1)$.

\subsection{Cluster size distribution}
\label{Cluster size distribution}

\begin{figure}
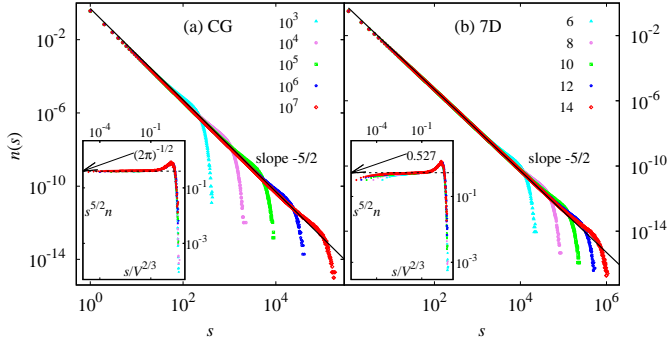

  \distribpFig
  \caption{Number density of complete clusters of size $s$ as a function of $s$ on the CG (a)  and  in 7D (b). 
           The inset shows $n(s) s^{5/2}$ vs.\ $s/V^{2/3}$ for the two systems.} 
  \label{Fig:distribpFig}
\end{figure}

\begin{figure}
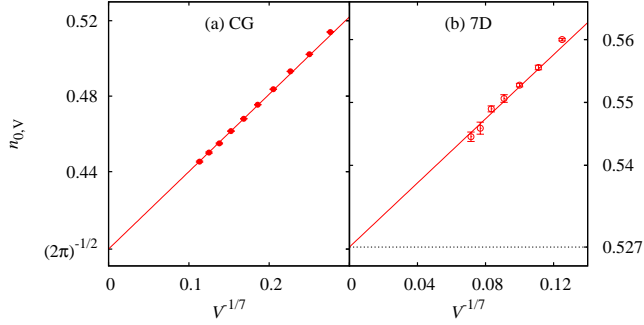

  \metricfaFig
  \caption{Metric factor $n_{0,V}$ vs.\ $V^{-1/7}$ on the CG (a) and in 7D (b). 
  The intercept respectively corresponds to values $(2\pi)^{-1/2}$ and $0.527$. }  
  \label{Fig:metricfaFig}
\end{figure}

Finally, we studied the size distribution of the complete, leaf-free and bridge-free clusters on the CG and in 7D. 

For complete configurations on the CG, Ref.~\cite{Krapivsky05} predicts that the critical number density $n(s,V)$ of clusters of size $s$ obeys a scaling form 
\begin{equation}
n(s,V) = n_0s^{-\tau} \tilde{n} (s/V^{d^*_{\rm f}}) \; , \hspace{5mm} (\tilde{n} (x \rightarrow 0) =1) 
\label{eq:size_dist}
\end{equation}
 with $n_0=(2\pi)^{-1/2}$ , volume fractal dimension $d^*_{\rm f}=2/3$ and exponent 
$\tau = 1+1/d^*_{\rm f}=5/2$. We numerically confirm this prediction, as shown in Fig.~\ref{Fig:distribpFig}(a), where the exponent $\tau = 5/2$ is represented 
by slope of the straight line. The inset of Fig.~\ref{Fig:distribpFig}(a) shows that $s^{5/2}n(s)$ vs.\ $s/V^{2/3}$ for different system volumes
collapses to a single curve which represents the scaling function $\tilde{n} (s/V^{2/3})$. 
Besides, according to~\cite{Krapivsky05} the scaling function $n(s,V)$ has the following extremal behaviors
\begin{equation}
 n(s,V) \simeq
 \left\{
  \begin{aligned}
            (2\pi)^{-1/2}s^{-5/2}   \  \hspace{1mm}  \ s \ll 1  \\ 
            {\exp}(-s^\gamma)       \  \hspace{1mm}  \ s \gg 1  \\ 
   \end{aligned}
   \right. 
   \label{Eq:scaling}
 \end{equation}
where $\gamma \cong 3$. Similar scaling behavior of $n(s,V)$ is observed for complete 
configurations in 7D, as shown in Fig.~\ref{Fig:distribpFig}(b).

In order to confirm the metric factor $n_0$ for the CG and determine $n_0$ in 7D, 
we count the total number $N_t$ of clusters of size $V^{4/7} < s < 2V^{4/7}$, which in theory scales as 
\begin{equation}
  N_t = n_0 \frac{2\sqrt2-1}{3\sqrt2} V^{1/7}[1+{\rm O}(V^{-y_2})]  
 \label{Eq:N_t}
\end{equation} 
where $y_2=4/7$ is the correction exponent of integration and summation.
The range $(V^{4/7}, 2V^{4/7})$ is chosen such that the lattice effect for small $s$ is suppressed 
while the universal function $\tilde{n}(x)$ in Eq.~(\ref{eq:size_dist}) has $x \sim {\rm O}(V^{-2/21}) \rightarrow 0$.
 We then define metric factor $n_{0,V} = 3\sqrt2 N_t V^{-1/7} /(2\sqrt2-1) $, which we fit by following ansatz
\begin{equation}
n_{0,V} = n_0+n_1 V^{-y_1}+n_2 V^{-y_2}  
  \label{Eq:n_0V}
\end{equation}

\begin{figure}
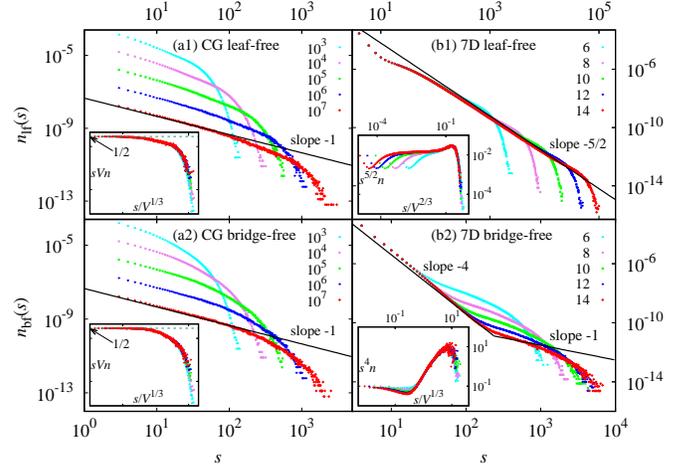

  \distribfFig
  \caption{Number density of (a1) leaf-free clusters on the CG, (a2) bridge-free clusters on the CG, (b1) leaf-free clusters in 7D, and (b2) bridge-free clusters in 7D 
           of size $s$ as a function of $s$. The inset shows (a1) $sVn_{\rm \ell f}(s)$ vs.\ $s/V^{1/3}$ on the CG,
           (a2) $sVn_{\rm b f}(s)$ vs.\ $s/V^{1/3}$ on the CG, 
          (b1) $n_{\rm \ell f}(s)s^{5/2}$ vs.\ $s/V^{2/3}$ in 7D, and (b2) $s^4n_{\rm b f}(s)$ vs.\  $s/V^{1/3}$ in 7D.} 
  \label{Fig:distribfFig}
\end{figure}

From the fits, we fix $y_1=1/7$, $y_2=4/7$ and estimate $n_0=0.3991(17)$, which is consistent with $(2\pi)^{-1/2} \approx 0.3989 $.
Fixing $n_0=(2\pi)^{-1/2}$ and $y_2=4/7$, we estimate $y_1=0.144(6) \approx 1/7$.
For 7D, we implement a similar procedure as for the CG. We find $n_0=0.527(7)$, 
which is not consistent with the CG value $(2\pi)^{-1/2}$. The resulting fits are summarized in Table~\ref{Tab:fitn0}. 
In Fig.~\ref{Fig:metricfaFig}, we plot $n_{0,V}$ of CG and 7D vs.\ $V^{-1/7}$,
providing further evidence that $n_0$ for CG tends to $(2\pi)^{-1/2}$ while $n_0$ for 7D tends to $\approx 0.527$.

\begin{table}[htbp]
  \begin{tabular}[t]{|l|l|l|l|l|l|}
     \hline
           &     $n_{0,V}$       & $n_0$        &  $y_1$       & $V_{\rm min}$ & DF/$\chi^2$ \\
     \hline
       {\multirow{4}{*}{CG}} & {\multirow{3}{*}{$$}}               & 0.3993(9)               &1/7         &$2^{15}$     &~5/4 \\
     \cline{3-6} 
     &                                                             & 0.3990(10)              &1/7         &$2^{16}$     &~4/4 \\     
     \cline{2-6}
      &                                                            & $(2\pi)^{-1/2}$         &0.142(2)    &$2^{15}$     &~5/4 \\     
     \cline{3-6}
      &                                                            & $(2\pi)^{-1/2}$         &0.144(4)    &$2^{16}$     &~4/4 \\     
     \cline{3-6}
     \hline 
       {\multirow{2}{*}{7D}} & {\multirow{2}{*}{$$}}               & 0.525(2)                &1/7         &$7^7$        &~5/5 \\
     \cline{3-6}
     &                                                             & 0.528(4)                &1/7         &$8^7$        &~4/5 \\  
     \cline{2-6}
     \hline          
  \end{tabular}
  \caption{Fit results of metric factor $n_{0,V}$ in Eq.~(\ref{Eq:n_0V}) for the CG and 7D.}
  \label{Tab:fitn0}
\end{table}

For leaf-free and bridge-free configurations on the CG, the results for the size distributions are shown in Figs.~\ref{Fig:distribfFig}(a1),(a2).  We find the behavior of both size
distribution are governed by a modified Fisher exponent $\tau^{\prime}=1$. To derive a scaling relation for the leaf-free and bridge-free configurations, 
consider the cluster distribution $n(s)$, which we assume satisfies 
\begin{equation}
  n(s,V) = n_0s^{-\tau^{\prime}}V^{-h} \tilde{n} (s/V^{d^*_\scrO}) \; , \hspace{3mm} (\tilde{n} (x \rightarrow 0) =1) 
  \label{Eq:scaling}
\end{equation} 
where $d^*_\scrO = d^*_{\ell {\rm f}} = d^*_{\rm bf}=1/3$ is 
the fractal dimension for both leaf-free and bridge-free configurations, while $n_0$, $\tau^{\prime}$ and $h$ are parameters to be determined. 
The occurrence of $V^{-h}$ reflects the fact that the linear parts in Fig.~\ref{Fig:distribfFig}(a1)-(a2) have a dependence upon the system volume. According to
the result that the total number of leaf-free and bridge-free clusters scales as $(1/6)\ln V$  at $p_c$, we perform the integral of equation (\ref{Eq:scaling})      
\begin{equation}
 \int_1^{V^{1/3}} n(s,V) ds \sim
 \left\{
  \begin{aligned}
            \frac{n_0}{3} (\ln V) V^{-h}                                        \ \hspace{1mm}  \text{if} \ \tau^{\prime} = 1     \\
            \frac{n_0}{-\tau^{\prime}+1} V^{\frac{1}{3}(-\tau^{\prime}+1)-h}    \  \hspace{1mm} \text{if} \ \tau^{\prime} \neq 1  \\ 
   \end{aligned}
   \right. 
   \label{Eq:integral}
 \end{equation}
combined  with 
\begin{equation}
  \int_1^{V^{1/3}} n(s,V) ds  =   \frac{\ln V}{6V} \\
  \label{Eq:number}
\end{equation} 
These two are only consistent in the case $\tau^{\prime} = 1$, and they also imply that $h = 1$. In the inset of Fig.~\ref{Fig:distribfFig}(a1)-(a2), we
plot $sVn(s,V)$ versus $s/V^{1/3}$ for both leaf-free and bridge-free clusters on the CG; the insets clearly demonstrate that our deduction that $n(s,V) = n_0s^{-1}V^{-1} 
\tilde{n} (s/V^{1/3})$ is correct. Besides, from equation (\ref{Eq:integral}) we can conjecture that $n_0=1/2$ which is also confirmed by the inset. The occurrence of the modified
Fisher exponent $\tau^{\prime}$ is related to the fact that cluster number is not proportional to the system volume.  This phenomenon can be also found in Ref.~\cite{Hao2016}. 

The cluster size distribution of leaf-free configurations in 7D follows the same behavior as the complete cluster configuration which should scale as $n_0s^{-5/2} \tilde{n} (s/V^{2/3})$.
However, when cluster size $s$ is very small such that $s \ll V^{2/3}$, the non-universal behavior of the scaling plot is very obvious, as shown in Fig.~\ref{Fig:distribfFig}(b1).
On the other hand, the  cluster size distribution of bridge-free configurations in 7D behaves differently and shows two-scaling behaviors. 
There are an extensive number of small clusters $s < V^{1/3}$, and the size distribution of these small clusters has a standard Fisher exponent $\tau= 1+1/d^*_{\rm bf}=4$
with $d^*_{\rm bf}=1/3$ the volume fractal dimension of the bridge-free clusters. 
When the cluster size is larger $s > V^{1/3}$,
the size distribution of these large clusters has the modified Fisher exponent $\tau^{\prime}=1$, which also governs the scaling behavior of the size distribution in the
CG case.

\begin{table*}
  \begin{center}  
  \begin{tabular}[t]{|l|l|l|l|l|l|l|l|} 
    \hline
         &\multicolumn{3}{c|}{CG}  & \multicolumn{3}{c|}{7D}   \\
    \hline
     $ $           & branch bonds                    & junction bonds              & non-bridge bonds            & branch bonds              & junction bonds             & non-bridge  bonds       \\
    \hline      
     $\rho_0$      & $0.999 999(3) \approx 1$        &$-0.000 000 1(2) \approx 0$  &$0.000 000 3(7) \approx 0$    & 0.985 330 4(6)           & 0.008 476 5(5)             & 0.006 193 1(7)          \\
    \hline 
     $y$           & $0.669(6) \approx 2/3$          &$0.661(6) \approx 2/3$       &$0.666(3) \approx 2/3$        & ~~~~~~~~~-               &$0.664(12)\approx 2/3$      &$0.671(17)\approx 2/3$   \\
    \hline 
    \vspace{-2mm}  & \multicolumn{3}{c|}{}  &  \multicolumn{3}{c|}{ }  \\  
   \hline 
     $ $           & complete cluster                & leaf-free cluster           & bridge-free cluster          & complete cluster         & leaf-free cluster  & bridge-free cluster \\
    \hline   
    $d^*_\scrO$    &$0.666 4(5)\approx 2/3$          &$0.333 7(17)\approx 1/3$     &$0.333 7(15)\approx 1/3$      &$0.669(9)\approx 2/3$     &$0.669(9)\approx 2/3$       &$0.332(7)\approx 1/3$        \\
    \hline 
    $a_0$          &$0.500 000 1(3) \approx 1/2$     &  ~~~~-                      &  ~~~~-                       & 0.450 278 06(9)          &  0.000 777 130(13)         & 0.000 975 139(27)           \\  
    \hline 
    $a_1$          &$0.167(9) \approx 1/6$           &$0.1672(9) \approx 1/6$      &$0.1673(10)\approx 1/6$       & 1.18(39)                 &  0.93(18)                  & ~~~~-                       \\
    \hline 
    $n_0$          &$0.3991(17)\approx(2\pi)^{-1/2}$ &  1/2                        & 1/2                          & 0.527(7)                 &  ~~~~-                     & ~~~~-                       \\
    \hline
  \end{tabular}
  \caption{Summary of estimated bond densities in the thermodynamic limit $\rho_0$ in Eq.~(\ref{Eq:Bond_Density}), 
           leading finite-size correction exponents for bond densities  $y$ in Eq.~(\ref{Eq:Bond_Density}); 
           volume fractal dimensions $d^*_\scrO$ in Eq.~(\ref{Eq:Fractal_dimensions}), 
           cluster number density $a_0$ of complete clusters in Eq.~(\ref{Eq:Number_Cluster2}) on the CG, 
           cluster number density $a_0$ of complete clusters, leaf-free clusters in Eq.~(\ref{Eq:Excess_number}) and bridge-free clusters in Eq.~(\ref{Eq:bf_density}) in 7D,
           coefficient of logarithmic term $a_1$  of complete clusters in Eq.~(\ref{Eq:Number_Cluster2}), leaf-free and bridge-free clusters $a_1$ in Eq.~(\ref{Eq:Number_Cluster3}) on the CG, 
           and excess cluster number $a_1$ of complete clusters and leaf-free clusters in Eq.~(\ref{Eq:Excess_number}) in 7D, 
           and metric factor $n_0$ in Eq.~(\ref{Eq:n_0V}).} 
  \label{Tab:summary}
  \end{center}
\end{table*}

\section{Discussion}
\label{Discussion}

We have studied the geometric structure of critical bond percolation on the complete graph (CG) and on the 7D hypercubic lattice with periodic boundary condition, 
by separating the occupied edges into three natural classes. We found that bridge-free clusters have the same volume fractal dimension ($1/3$) on the CG and in 7D 
while leaf-free configurations do not (1/3 and 2/3, respectively). 
This observation answered the question raised in the section~\ref{Introduction} 
whether the mean-field theory always holds as a predictor of 
all kinds of exponents governing critical behavior above the upper critical dimension. 
Obviously, the answer is no.

The study of three kinds of bond densities on the CG and in 7D  provided more details about the geometric properties of percolation between the two 
models. Similar to the 2D case~\cite{XuWangZhouTimDeng2014}, the density of branches in 7D is only very weakly dependent on the system size although 
they occupy around 98.5 percent of the occupied bonds in the system. On the other hand the density of branches on the CG tends to 1 in the thermodynamic limit. 
The different behaviors of density of branches between the CG and 7D may result in the difference of leaf-free cluster fractal dimensions between CG and 7D.

From our work, we obtain the following general picture of percolation on the CG and in 7D:

On the CG, in the limit of $V \to \infty$, the connectivity becomes identical to a Bethe lattice with an infinite number of possible bonds at each vertex, but with on the average just one of those bonds being occupied at the critical point.  The number of blobs (bridge-free clusters) $N_{\rm bf}$ is just a few, so their density $N_{\rm bf}/V = \rho_{\rm bf}$ goes to zero. 
These blobs are most likely in the giant clusters, which have a size of O($V^{2/3}$); 
most of the rest of the clusters are too small to have any loops.
The giant clusters and the few blobs are what distinguishes the CG from the Bethe lattice, 
which is problematic here because of its large surface area.  
The critical exponents such as $\tau = 5/2$ are the same for the CG and the Bethe lattice, being the mean-field values, but other properties are different.

In 7D percolation, the critical exponents are also mean-field, so in that sense the system is similar to the CG and Bethe lattice.  The 7D system does have blobs like the giant components of the CG, 
with a similar size $\sim V^{2/3}$, however in 7D the blobs are much more numerous and represent a finite fraction of the clusters in the system.  Still, on an overall scale, the collection of clusters has a tree structure, decorated with blobs in various places.  The tree can have branch points where more than two junction bonds visit a single point, or where more than two junction bonds connect to a blob.

Finally, we compare these results with the 2D results~\cite{XuWangZhouTimDeng2014},
 which is below the critical dimension 6, so universality holds. 
The general scenario for the geometric structure of critical percolation clusters is the same for all finite dimensions: the leaf-free clusters have the same
fractal dimension as the original percolation cluster, and the bridge-free clusters (blobs) have the dimension of backbone clusters. 
For $d \geq d_u$, the blobs are mostly unicycles, while the blobs in lower dimensions have many cycles. It would be of interest to check this scenario in more detail.

Table~\ref{Tab:summary} summarizes the estimates presented in this work, including our conjectures for exact values for several of the quantities.

A natural question to ask is to what extent the results of percolation on a seven-dimensional lattice with periodic boundary carry over to the percolation on a seven-dimensional lattice 
with free boundary conditions. Although the largest clusters with free boundary conditions have fractal dimension $d_{\rm f}=4$ which is independent of spatial dimension, 
at the pseudocritical point with free boundary conditions largest clusters can have fractal dimension $d_{\rm f}=2d/3$~\cite{kennaBerche2016}. It is of interest to see what will 
happen for leaf-free and bridge-free clusters and relative geometric properties at the critical point and pseudocritical point in 7D with free boundary conditions. 
Studies in higher dimensions would also be interesting although there will be difficulties due to the limitations on the size of the system that can be simulated.

\section{Acknowledgments}
We acknowledge the contribution of P. J. Zhu to the algorithm for classifying bonds. 
We also want to thank Y. B. Zhang for his help in making Fig.~\ref{Fig:cartoonfFig}.  
This work was supported by the National Natural Science Fund for Distinguished 
Young Scholars (NSFDYS) under Grant No.~11625522 (Y.J.D), the National Natural 
Science Foundation of China (NSFC) under Grant No. and~11405039 (J.F.W), 
and the Fundamental Research Fund for the Central Universities under 
Grant No.~J2014HGBZ0124 (J.F.W).  R.M.Z. thanks the hospitality of the UTSC while this paper was written.


\end{document}